# Design of an indium arsenide cell

# for near-field thermophotovoltaic devices

Daniel Milovich[a], Juan Villa[2], Elisa Antolin[b], Alejandro Datas[b], Antonio Marti[b], Rodolphe Vaillon[c,d,b], Mathieu Francoeur[a,†]

[a]Radiative Energy Transfer Lab, Department of Mechanical Engineering, University of Utah, Salt Lake City, UT 84112, USA

[b]Instituto de Energía Solar, Universidad Politécnica de Madrid, 28040 Madrid, Spain

[c]IES, Univ Montpellier, CNRS, Montpellier, France

[d]Univ Lyon, CNRS, INSA-Lyon, Université Claude Bernard Lyon 1, CETHIL UMR5008, F-69621, Villeurbanne, France

**Abstract**

An indium arsenide photovoltaic cell with gold front contacts is designed for use in a near-field thermophotovoltaic (NF-TPV) device consisting of millimeter-size surfaces separated by a nanosize vacuum gap. The device operates with a doped silicon radiator maintained at a temperature of 800 K. The architecture of the photovoltaic cell, including the emitter and base thicknesses, the doping level of the base, and the front contact grid parameters, are optimized for maximizing NF-TPV power output. This is accomplished by solving radiation and charge transport in the cell via fluctuational electrodynamics and the minority charge carrier continuity equations, in addition to accounting for the shading losses due to the front contacts and additional series resistance losses introduced by the front contacts and the substrate. The results reveal that these additional loss mechanisms negatively affect NF-TPV performance in a non-negligible manner, and that the maximum power output is a trade-off between shading losses and series resistance losses introduced by the front contacts. For instance, when the cell is optimized for a $1 \times 1$ mm$^2$ device operating at a vacuum gap of 100 nm, the losses introduced by the front contacts reduce the maximum power output by a factor of $\sim 2.5$ compared to the idealized case when no front contact grid is present. If the optimized grid for the $1 \times 1$ mm$^2$ device is scaled up for a $5 \times 5$ mm$^2$ device, the maximum power output is only increased by a factor of $\sim 1.08$ with respect to the $1 \times 1$ mm$^2$ case despite an increase of the surface area by a factor of 25. This work demonstrates that the photovoltaic cell in a NF-TPV device must be designed not only for a specific radiator temperature, but also for specific gap thickness and device surface area.

[†] Corresponding author.
*Email address:* mfrancoeur@mech.utah.edu

## 1. Introduction

Thermophotovoltaic (TPV) power generators convert infrared photons from a terrestrial heat source into electricity and consist of a high-temperature radiator and a narrow-bandgap photovoltaic (PV) cell separated by an infrared-transparent gap. TPV devices have been proposed for use in several applications as they offer flexibility in terms of the heat source employed to maintain the radiator at a high temperature, including nuclear fuels, waste heat, and concentrated solar radiation, amongst others [1]. For instance, solar TPV converters are an attractive alternative to traditional solar PV due to their maximum overall theoretical efficiency of 85.4% [2]. Traditional TPV devices operate in the far-field regime of thermal radiation in which the gap separating the radiator from the cell is larger than the characteristic thermal wavelength emitted. In this regime, radiative transfer is mediated by propagating waves and is therefore limited by Planck's blackbody distribution. In the near-field regime of thermal radiation, which refers to the case where objects are separated by sub-wavelength gaps, radiation exchange can exceed the blackbody limit owing to tunneling of evanescent waves that are confined within a distance of a wavelength, or less, normal to the surface of a heat source [3]. To potentially improve TPV performance, Whale and Cravalho [4] proposed near-field TPV (NF-TPV) devices capitalizing on evanescent waves by separating the radiator and cell by a sub-wavelength vacuum gap.

Despite more than a decade of active research, a handful of laboratory-scale experiments have reported modest NF-TPV performances. DiMatteo et al. [5] conducted experiments using a silicon (Si) radiator and an indium arsenide (InAs) cell in which the vacuum gap with a nominal size of 1 μm was modulated via a piezoactuator. The near-field enhancement was interpreted from an in-phase increase in short-circuit current with the oscillating vacuum gap. Fiorino et al. [6] characterized NF-TPV performance enhancement in a nanopositioning platform using a commercial indium arsenide antimonide (InAsSb) cell and a microsize Si radiator. While this work showed a ~ 40-fold enhancement in power output relative to the far-field limit for a vacuum gap thickness of 60 nm and a radiator temperature of 655 K, an estimated efficiency of ~ 0.02% and a maximum power output on the order of a few tens of nW were reported, since only ~ 5.6% of the cell active area was illuminated. Inoue et al. [7] fabricated and characterized a one-chip NF-TPV device made of a thin film Si radiator, an indium gallium arsenide (InGaAs) cell, and an undoped Si layer coated on the cell. For a temperature difference of ~ 700 K between the radiator and cell and an average vacuum gap of 140 nm, a 10-fold enhancement of the photocurrent over the far-field limit was reported, while the estimated conversion efficiency was limited to 0.98%. Bhatt et al. [8] proposed a NF-TPV platform relying on an integrated nano-electromechanical system enabling modulation of the vacuum gap separating the germanium PV cell and the radiator made of a thin chrome film on tungsten. For radiator and cell temperatures of respectively 880 K and 300 K, the power output increased by a factor of 11 by reducing the vacuum gap from 500 nm down to ~ 100 nm; the conversion efficiency was not reported.

One of the major barriers preventing the physical realization of NF-TPV devices beyond laboratory-scale proof-of-concepts is the ability to maintain a nanosize vacuum gap between macroscale surfaces. However, recently, the ability to maintain vacuum gaps on the order of 100 to 200 nm, distances that are small enough to obtain substantial near-field radiative transfer enhancement, between millimeter to centimeter-sized surfaces have been demonstrated [9-12]. The other major barrier to NF-TPV device implementation is the lack of narrow-bandgap PV cells designed specifically for operation under near-field illumination. Traditional cell front contacts consisting of a metallic grid and busbars pose unique challenges with NF-TPV as the dimensions of these contacts are on the order of or larger than the vacuum gap distance required to observe significant near-field enhancement. In addition, the vacuum gap thickness in NF-TPV is analogous to the concentration factor in concentrated PV (CPV). In the near field, the amount of illumination seen by the cell is "magnified" via evanescent waves rather than by mirrors or lenses in CPV. Similar to CPV, the large amount of illumination in NF-TPV is expected to result in significant series resistance losses [13]. Despite several NF-TPV theoretical works [4,14-40], few have addressed the issue of the front contacts. Of these, some have proposed the use of thin layers of transparent, conducting materials, such as transparent conducting oxides [32,33,40]. The additional series resistance and shading losses introduced by the front contacts were neglected in Refs. [32,33], while the impact of the front contacts on the cell performance was considered only through shading losses in Ref. [40]. Recent works on thermionic-enhanced NF-TPV devices [41,42] included the effect of additional series resistance losses introduced from charge carrier extraction, but one advantage of these hybrid devices is that the use of a metallic front contact grid can be avoided. Microsize cells made of indium antimonide (InSb) were designed by accounting for series resistance losses and fabricated for use in a NF-TPV laboratory-scale experimental bench operating at cryogenic cell temperature ($\sim 77$ K) [43,44]. Using a graphite microsphere radiator at a temperature of $\sim 800$ K, a power density three orders of magnitude larger than the previous state-of-the-art [6-8] was measured, and a record NF-TPV conversion efficiency of 14% was reported for a sub-100 nm vacuum gap [45]. This conversion efficiency largely exceeds the previously reported maximum value of 0.98% [7], thus showing the importance of designing PV cells for operation under near-field illumination. However, the InSb cells discussed in Refs. [43-45] cannot be readily applied to high-temperature devices made of macroscale surfaces.

The objective of this work is therefore to design a narrow-bandgap InAs photovoltaic cell with gold (Au) front contacts suitable for a NF-TPV device consisting of macroscale surfaces ($\sim$ mm$^2$) separated by a nanosize vacuum gap. This is done by determining the cell architecture and doping level, as well as the front contact grid parameters, maximizing NF-TPV power output while accounting for shading and series resistance losses introduced by the front contacts. Solutions of radiation and charge transport via fluctuational electrodynamics and the continuity equations for minority charge carriers clearly show that a cell design that includes the front contact grid must account not only for the radiator temperature, but also for the gap thickness and size of the surfaces.

The rest of the paper is organized as follows. Section 2 is devoted to the description of the problem, with a specific focus on the optical and electrical properties needed to perform radiation and charge transport calculations. Next, the results of the InAs cell design for 1 × 1 mm$^2$ and 5 × 5 mm$^2$ NF-TPV devices are discussed. Finally, concluding remarks are provided.

## 2. Methodology

### 2.1. Problem description

The NF-TPV device analyzed in this work is shown in Fig. 1(a) and consists of a p-doped Si radiator and a PV cell made of InAs (p- on n- on n-substrate configuration) separated by a vacuum gap of thickness $d$. Both the radiator and cell are characterized by square macroscale surfaces of $1 \times 1$ mm$^2$ or $5 \times 5$ mm$^2$.

The radiator doping level, $N_{a,radiator}$, is fixed at $10^{19}$ cm$^{-3}$ because the real part of the refractive index of p-doped Si is nearly identical to that of p-doped InAs in the spectral band of interest, thus maximizing radiation transfer [46]. A radiator temperature of 800 K is assumed since it is the maximum value that has been achieved in experiments so far [45], while the temperature of the InAs cell is fixed at 300 K.

InAs is selected as the semiconductor material for the PV cell because it has a narrow bandgap at 300 K (0.354 eV [47]), and it does not require to be cooled down to low temperatures to operate like InSb [43-45]. The InAs cell is designed for fabrication by molecular beam epitaxy, which constrains its architecture. A thickness $t_{sub}$ of 500 μm and doping level, $N_{d,sub}$, of $1.5 \times 10^{17}$ cm$^{-3}$ for the substrate are selected based on typical values of commercially available n-doped InAs wafers [48]. The substrate is assumed to be inactive with respect to photocurrent generation in comparison to the epitaxially grown layers. This assumption looks at the worst case scenario in which the high recombination rate at the substrate/epitaxial-layer interface avoids collecting the carriers that are photogenerated carriers in the substrate. The total thickness of the p-n junction, $t_p + t_n$, is fixed at 3 μm where the top, p-doped layer is hereafter referred to as the emitter, and the bottom, n-doped layer is called the base. Within the 3 μm total thickness, the thicknesses of the emitter, $t_p$, and base, $t_n$, are allowed to vary between 0.40 and 2.60 μm. The doping level of the emitter is limited due to diffusion mechanisms. Beryllium is used as the p-dopant, which is known to diffuse toward the base [49]. Thus, a non-uniform spatial doping profile will be established. By fixing the doping level of the emitter, $N_{a,emitter}$, to $10^{18}$ cm$^{-3}$, a compromise is made between selecting a doping level as high as technically possible in order to create a large diffusion potential while also preventing the creation of a significant spatial doping profile. As a result, the doping level of the emitter in the simulations is assumed to be uniform. Four possible values are selected for the doping level of the base, $N_{d,base}$, namely $10^{16}$, $10^{17}$, $10^{18}$, and $10^{19}$ cm$^{-3}$.

The cell front contacts are typically composed of Au/Zn/Au for p-type contacts [50]. As the Zn layer is very thin with respect to Au, the front and back contacts are modeled as evaporated Au with uniform resistivity, $\rho_c$, of $10^{-7}$ $\Omega$m; this simplification does not affect the main conclusions of this paper. The back Au contact has the additional benefit of acting as an optical back reflector. Both the front and back contacts have fixed thickness, $t_c$, of 200 nm. The front contact grid consists of the busbar and fingers having lengths equal to the lateral dimension, $a$, of the cell (see Fig. 1(b)). The free parameters for the front contact grid include the width of the busbar, $w_{bb}$ (20 to 1500 µm), the width of the fingers, $w_f$ (20 to 1500 µm), and the number of fingers, $n_f$ (1 to 200). Note that a choice of $n_f$ implies the choice of $l$, the spacing between the fingers, assuming they are equally spaced.

A minimum vacuum gap of 100 nm is chosen based on current state-of-the-art fabrication of near-field radiative heat transfer devices with millimeter-size surfaces [11]. The vacuum gap $d$ is defined as the distance between the bottom surface of the radiator and the top surface of the p-doped InAs layer (see Fig. 1(a)). Maintaining a vacuum gap distance smaller than the thickness of the front contact grid is possible by etching a series of trenches in the Si radiator in positions corresponding to those of the fingers and busbar. For simplicity, it is assumed that the radiator covers the entire inter-finger spacing. While this assumption leads to a small underestimation of the shading losses, it does not affect the main conclusions of this work. The fixed and free parameters for the InAs cell and the front contact grid are summarized in Table 1.

## 2.2. Radiation and charge transport in the cell

One-dimensional radiation and charge transport along the $z$-direction is considered. Two-dimensional effects are negligible because the vacuum gap thickness is much smaller than the radiator and cell surface dimensions. In all simulations, the emitter and base are discretized into control volumes, which allow for calculation of the $z$-spatial distribution of absorbed radiation and carrier generation in the p-n junction as well as the total radiative power absorbed by the p-n junction and the cell (p-n junction, substrate and back contacts). Converged results are obtained by discretizing the emitter and the base into a total of 900 uniform control volumes, leading to individual control volumes having an equal thickness of 3.33 nm. The specific number of control volumes in the emitter and the base depends on their thickness.

Radiation transport is calculated assuming flat surfaces. Since the width of the fingers, $w_f$, and their separation distance, $l$, are sufficiently large in comparison to the characteristic wavelength of thermal emission ($\sim 3.6 \, \mu m$), any radiation scattering due to the front contact is assumed to be negligible. Under this assumption, the front contact grid is not included directly in the radiation transport model. Instead, its presence is taken into account via a shading effect when calculating the photocurrent generated by the cell (see Section 2.2.2).

### 2.2.1. Radiation transport and optical properties

Radiation transport is calculated using fluctuational electrodynamics, where a fluctuating current representing thermal emission is added to Maxwell's equations [51]. The expression for computing the radiation absorbed by the control volumes in the cell, provided in Refs. [18,52], requires the dielectric function of all layers shown in Fig. 1(a) as inputs. The specific parameters needed to calculate the dielectric functions discussed hereafter are provided in Table 2. Note that luminescent photons and thus photon recycling are neglected; this simplification does not affect the main conclusions of this work.

The dielectric function of p-doped Si ($10^{19}$ cm$^{-3}$) and Au are calculated using a Drude model [53,54]:

$$\varepsilon(\omega) = \varepsilon_\infty - \frac{\omega_p^2}{\omega(\omega + i\Gamma)} \tag{1}$$

where $\varepsilon_\infty$ is the high-frequency permittivity, $\omega_p$ is the plasma frequency, and $\Gamma$ is the damping coefficient due to free carriers.

The optical properties of InAs (both p- and n-doped) are modeled by taking into account absorption by the lattice (phonons), absorption by the free carriers, and interband absorption. The lattice and free carrier contribution to the dielectric function of InAs, $\varepsilon_{FCL}$, is calculated using a Drude-Lorentz model [55-57]:

$$\varepsilon_{FCL}(\omega) = \varepsilon_\infty \left( 1 + \frac{\omega_{LO}^2 - \omega_{TO}^2}{\omega_{TO}^2 - \omega^2 - i\omega\gamma} - \frac{\omega_p^2}{\omega(\omega + i\Gamma)} \right) \tag{2}$$

where $\omega_{LO}$ and $\omega_{TO}$ are the longitudinal and transverse optical phonon frequencies, while $\gamma$ is the damping coefficient due to phonons. Radiation absorption by the lattice and free carriers is significant only at frequencies below the cell absorption bandgap.

The dielectric function for interband transitions, $\varepsilon_{IB}$ (= $\varepsilon'_{IB} + i\varepsilon''_{IB}$) is calculated from the complex refractive index for interband transitions, $m_{IB}$ (= $m'_{IB} + im''_{IB}$), using the relations $\varepsilon'_{IB} = (m'_{IB})^2 - (m''_{IB})^2$ and $\varepsilon''_{IB} = 2m'_{IB}m''_{IB}$. The imaginary part of the refractive index is calculated directly from the interband absorption coefficient, $\alpha_{IB}$, as follows:

$$m''_{IB}(\omega) = \frac{\alpha_{IB}c_0}{2\omega} \tag{3}$$

where $c_0$ is the speed of light in vacuum. Once $m''_{IB}$ is known, the real part of the refractive index, $m'_{IB}$, can be constructed using the Kramers-Krönig relations [60].

Due to the narrow bandgap of InAs, energy states near the bottom of the conduction band are easily filled, and the semiconductor can become degenerate beyond a certain doping concentration. At high enough doping concentration, the Moss-Burstein shift plays a decisive role in the behavior of the interband absorption coefficient [61-63]. When a semiconductor is n-doped degenerately, the Fermi level lies above the conduction band edge. Thus, for an interband transition to occur, a photon must have a minimum energy equal to the bandgap energy plus the difference in energy between the conduction band edge and the lowest unfilled energy state in the conduction band. As a result, interband absorption in InAs is shifted to higher energies as doping concentration increases due to the Moss-Burstein shift. The equations for calculating the interband absorption coefficient, $\alpha_{IB}$, accounting for the Moss-Burstein shift are provided in Ref [64].

The interband absorption coefficient and the dielectric function for interband transitions are calculated only for frequencies greater than the angular frequency corresponding to the bandgap of InAs ($\omega_g = 5.38 \times 10^{14}$ rad/s), whereas $\varepsilon_{FCL}$ is calculated at all frequencies in the spectral band of interest. Lastly, the complete dielectric function of InAs, accounting for all three absorption mechanisms, is given by $\varepsilon(\omega) = \varepsilon_{FCL}(\omega) + \varepsilon_{IB}(\omega)$.

### 2.2.2. Electrical transport and electrical properties

Radiation absorbed by the control volumes in the cell is used to calculate the local generation rate of electron-hole pairs, which in turn serves as an input to the continuity equations for minority charge carriers under the low-injection approximation [18]. Solution of the continuity equations for minority charge carriers with and without illumination enables calculation of the photocurrent, dark current, current-voltage characteristics, electrical power output and conversion efficiency. The effect of series resistance losses introduced by transverse current flow through the emitter and base is taken into account when solving the continuity equations for minority charge carriers. Note that non-degenerate conditions are assumed in the electrical transport model.

The properties required to perform the electrical transport calculations are the mobility of majority and minority carriers in the emitter, base, and substrate, in addition to minority carrier lifetimes and surface recombination velocity. An empirical Caughey-Thomas-like model [65] is used to compute mobility, $\mu$, as a function of doping concentration and temperature for both majority and minority charge carriers:

$$\mu(N,T) = \mu_{\min} + \frac{\mu_{\max}\left(\dfrac{300}{T}\right)^{\theta_1} - \mu_{\min}}{1 + \left(\dfrac{N}{N_{ref}\left(\dfrac{T}{300}\right)^{\theta_2}}\right)^{\phi}} \tag{4}$$

All parameters required for the Caughey-Thomas model as well as the models and parameters used for minority carrier lifetime due to radiative recombination, Auger recombination, Shockley-Read-Hall recombination, and surface recombination velocity are provided in Table 3. The total minority carrier lifetimes are calculated using Matthiessen's rule.

The results obtained from solving the minority charge carrier continuity equations must be modified to account for the presence of the front contact grid (additional series resistance and shading losses) and the substrate (additional series resistance). These contributions are included in the model following the method of Ref. [66], which accounts for the additional series resistance losses by calculating the Joule losses introduced by current flowing through each component of the device, namely the transverse flow of current through the substrate, the lateral flow of current through the emitter, and current flow through the grid (fingers and busbar). Also, note that contact resistance between the Au front contact grid and the emitter is not considered. Experimental measurements of the contact resistance of the interface between these two materials are available but are application-specific with significant variation depending on the composition of the metallic contact and the InAs layer [67,68]. In practice, this resistance would be measured for the specific materials to be used in the fabricated device and could then be added to the grid optimization model. The series resistance contributions of the substrate ($r_{sub}$), emitter (due to lateral current flow) ($r_e$), busbar ($r_{bb}$), and grid fingers ($r_f$) normalized to unit area ($\Omega m^2$) are calculated as follows [66]:

$$r_{sub} = \rho_{sub} t_{sub} \tag{5}$$

$$r_e = \frac{1}{12} l^2 R_{sh} \tag{6}$$

$$r_{bb} = \frac{1}{12} a n_f^2 l^2 \frac{\rho_c}{w_{bb} t_c} \tag{7}$$

$$r_f = \frac{1}{3} a^2 l \frac{\rho_c}{w_f t_c} \tag{8}$$

where $\rho_{sub}$ is the resistivity of the substrate ($\Omega$m), and $R_{sh}$ is the sheet resistance of the emitter ($\Omega$) calculated using the resistivity of the emitter $\rho_e$ ($\Omega$m). The resistivity of the substrate and emitter, and the sheet resistance are respectively given by:

$$\rho_{sub,e} = \frac{1}{\mu_{maj} N e} \tag{9}$$

$$R_{sh} = \frac{\rho_e}{t_p} = \frac{1}{\mu_h N_a e} \frac{1}{t_p} \tag{10}$$

where $\mu_{maj}$ is the mobility of majority carriers (electrons in the substrate, $\mu_e$, holes in the emitter, $\mu_h$), $N$ is the doping level of the substrate or emitter ($N_d$ for electrons in the substrate, $N_a$ for holes in the emitter), and $e$ is the fundamental charge. The total additional series resistance normalized to unit area, $r_s$, introduced by the grid and the substrate is the sum of Eqs. (5) to (8).

The grid is optimized to extract maximum power from the cell, which introduces a trade-off between the additional series resistance, $r_s$, and shading losses. Their impact on the current-voltage characteristics is considered by introducing a voltage loss:

$$V' = V - I' \frac{r_s}{A_{cell}} \tag{11}$$

where $V'$ is the modified voltage, $V$ is the applied voltage, $I'$ is the net total current flowing through the device (i.e., photocurrent minus dark current) modified due to the shading effects of the grid, and $r_s/A_{cell}$ is the additional series resistance of the grid and substrate in units of $\Omega$. Shading effects are accounted for geometrically, meaning that the fraction of the total area that is shaded by the grid is removed from the photocurrent:

$$I' = \left[ J_{ph}(1 - F_s) - J_{dark} \right] A_{cell} \tag{12}$$

where $J_{dark}$ and $J_{ph}$ are the dark current density and the photocurrent density, respectively, calculated from solving the minority charge carrier diffusion equations without any modifications (i.e., assuming the entire cell area is illuminated), and $F_s$ is the fraction of the total cell surface area that is shaded by the grid.

The maximum power output, $P_{mpp}$, in units of W is calculated as:

$$P_{mpp} = I'_{mpp} V'_{mpp} \tag{13}$$

where $I'_{mpp}$ and $V'_{mpp}$ represent the current and voltage at the maximum power point, respectively, of the modified current-voltage characteristics. The p-n junction efficiency, $\eta_{jun}$, and cell efficiency, $\eta_{cell}$, which include the p-n junction, substrate and back contacts, are defined as:

$$\eta_{jun} = \frac{P_{mpp}}{P_{abs,jun}} \tag{14}$$

$$\eta_{cell} = \frac{P_{mpp}}{P_{abs,cell}} \tag{15}$$

where $P_{abs,jun}$ and $P_{abs,cell}$ are the total radiative power absorbed by the p-n junction and the cell, respectively, determined from the radiation transport model. The efficiencies are defined in this way because it is not possible to calculate the incident radiation power when evanescent waves are accounted for [46].

## 3. Results and discussion

### 3.1. Selection of the cell parameters: emitter and base thickness, and doping level of the base

Fig. 2 shows the total absorption coefficient of InAs (lattice, free carrier, and interband) as a function of the doping level of the base, while Fig. 3 provides the spectral-spatial absorption distribution within the cell ($d = 100$ nm) for the lowest doping level considered, $10^{16}$ cm$^{-3}$ (panel (a)), and for the highest doping level considered, $10^{19}$ cm$^{-3}$ (panel (b)). Note that validation of the calculation of the InAs absorption coefficient is presented in Fig. S1. The cell shown in Fig. 3 uses an optimal set of parameters leading to the highest electrical power output at $t_p = 0.40$ μm, $t_n = 2.60$ μm, and $N_{d,base} = 10^{16}$ cm$^{-3}$. The junction between the emitter and the base is marked in Fig. 3 by a vertical dashed line.

The optimal set of parameters can be understood in terms of the physics of near-field thermal radiation and the properties of InAs. Unlike propagating electromagnetic waves, evanescent waves have a small penetration depth, ranging from approximately a wavelength down to the vacuum gap thickness $d$ for materials supporting surface plasmon polaritons, such as doped Si [69]. Additionally, n-doped InAs exhibits longer minority carrier lifetimes in comparison to p-doped InAs [56]. In Fig. 3(a), recombination in the highly p-doped emitter is Auger-limited while recombination in the lower-doped base is limited by radiative recombination [70]. Consequently, it is advantageous to minimize the emitter thickness $t_p$ such that the base absorbs as much of the incident radiation as possible.

As the doping concentration of the base increases, absorption below the bandgap due to the lattice and free carriers increases by orders of magnitude (see Fig. 2). In addition, for the largest doping concentration ($N_{d,base} = 10^{19}$ cm$^{-3}$), absorption above the bandgap is shifted to higher frequencies (energies) due to the Moss-Burstein shift. These two phenomena have negative impacts on the cell performance. Increased absorption below the bandgap due to free carriers and the lattice does not produce electron-hole pairs and thus contributes to diminishing conversion efficiencies [71]. The shift of the bandgap to higher energies negatively affects the device performance due to the relatively low temperature of the radiator (800 K). For a relatively high doping concentration of $10^{19}$ cm$^{-3}$, absorption below the bandgap increases substantially while absorption near the bandgap is reduced compared to a relatively low negative doping concentration of $10^{16}$ cm$^{-3}$; this effect is distinctly visible in the spectral-spatial absorption distributions in Fig. 3. This explains the choice of $10^{16}$ cm$^{-3}$ for the doping concentration of the base. While a higher doping concentration would yield a larger built-in voltage and a larger depletion region, these effects are outweighed by the impact of the Moss-Burstein shift on absorption within the cell.

### 3.2. Device performance with optimized front contact grid parameters

Fig. 4 shows the radiative flux absorbed by the cell (without shading losses) for gap thicknesses ranging from 100 nm up to 10 μm (far-field limit). For comparison, the flux between two blackbodies maintained at temperatures of 800 K and 300 K is also plotted. For vacuum gap thicknesses smaller than 1 μm (marked by the vertical dashed line in Fig.

4), radiation transfer is enhanced above the blackbody limit. The shaded region to the left of the vertical dashed line at $d = 1$ μm shows the extraneous energy that can be harvested in the near field beyond the blackbody limit. At a vacuum gap thickness of 100 nm, the radiation absorbed by the cell exceeds the blackbody limit by a factor of ∼ 5 and the far-field value by a factor of ∼ 13. In this case, the total radiation absorbed is equal to 117,950 Wm$^{-2}$, the equivalent of ∼ 118 suns.

Table 4 provides the parameters of the front contact grid maximizing the power output for the $1 \times 1$ mm$^2$ and $5 \times 5$ mm$^2$ devices operating with gap thicknesses of 10 μm and 100 nm along with the corresponding maximum power output, $P_{mpp}$, the p-n junction efficiency, $\eta_{jun}$, and the cell efficiency, $\eta_{cell}$, when the additional losses introduced by the front contact grid and substrate are and are not included. In the following, "without additional losses" is defined as the absence of both shading and additional series resistance losses $r_s$ (sum of Eqs. (5) to (8)), while "with all losses" refers to the inclusion of both of these losses. The maximum power output with the optimized grid parameters is also shown in Fig. 5 with all losses and without additional losses for a $1 \times 1$ mm$^2$ device with vacuum gap thicknesses ranging from 100 nm to 10 μm. Fig. S2 provides the optimized front contact grid parameters as a function of the gap thickness (100 nm to 10 μm) for the $1 \times 1$ mm$^2$ device with all losses.

The optimal front contact grid architecture is a strong function of the vacuum gap thickness and the dimensions of the device. For example, a $1 \times 1$ mm$^2$ device operating in the far field requires a busbar with a width of 31 μm and 6 fingers having widths of 20 μm to maximize power output. The same device operating with a gap thickness of 100 nm requires a thicker busbar (77 μm) and more fingers (13 fingers having widths of 20 μm) which are more closely spaced together to extract charge carriers with minimal losses due to the increase in illumination from evanescent waves. As a result, the shading losses increase, but this is outweighed by the reduction in additional series resistance losses $r_s$. The maximum power output for the far-field case is 0.0690 mW (1x1 mm$^2$ device), while it is 2.25 mW for the near-field case when no additional losses are considered, resulting in an enhancement factor of ∼ 33. When all losses are taken into account, the maximum power output is 0.0435 mW and 0.910 mW for the far-field and the near-field cases, respectively, such that the corresponding near-field enhancement is reduced to a factor of ∼ 21. Therefore, similar to CPV, it is clear that series resistance losses are more severe in the near field due to the substantial level of illumination. Note that the power enhancement is larger than the flux enhancement since the efficiency of the 100-nm-thick gap NF-TPV device is larger than the efficiency of the far-field device (see Table 4).

If the optimized grid for the $1 \times 1$ mm$^2$ device is simply scaled up for the $5 \times 5$ mm$^2$ device (i.e., all grid parameters are held constant except for the number of fingers and the finger length equal to the lateral dimension of the cell), the maximum power output is 0.982 mW when all losses are included, which is only ∼ 1.08 times larger than that of the $1 \times 1$ mm$^2$ device despite an increase of the surface area by a factor of 25 (see Table 4). If the grid is optimized specifically for the $5 \times 5$ mm$^2$ device, the maximum power output increases to 3.53 mW when all loss mechanisms are included, which is ∼ 3.6 times larger than for the case when the grid is scaled up from the $1 \times 1$ mm$^2$ device; the resulting near-field enhancement is ∼ 8.9. Clearly, the optimized grid parameters for the $1 \times 1$ mm$^2$ device do not scale to larger surface area devices. The front contact grid must be designed specifically for the dimensions of a given device in addition to the vacuum gap distance at which the device operates. Also, the near-field enhancement of power output decreases with increasing the cell surface area when all losses are considered, while the near-field enhancement is essentially the same (~ 33) for both the 1 × 1 mm$^2$ and 5 × 5 mm$^2$ devices when the additional losses are neglected. This suggests that the cell suffers from scalability issues that could potentially be resolved by relaxing some of the parameters fixed in the design process. Note that even though the choice of a thin, 0.4-μm-thick emitter yields a large sheet resistance, the contribution due to the fingers is the largest for both the 1 × 1 mm$^2$ and the 5 × 5 mm$^2$ devices.

Fig. 6 shows the current-voltage characteristics of the 1 × 1 mm$^2$ device ($d$ = 100 nm and 10 μm) under three conditions, namely without additional losses, when only the additional shading losses introduced by the front contact grid are considered, and when all losses are accounted for. The current-voltage characteristics are the superposition of the current-voltage characteristics when the cell is in dark conditions and when it is subject to illumination. As it can be observed, the output voltage never exceeds ~ 0.125 V. Assuming an intrinsic concentration of $n_i$ = 10$^{15}$ cm$^{-3}$ for InAs at 300 K, we find that $n_i \exp(eV / 2k_b T) \approx$ 10$^{16}$ cm$^{-3}$ ≲ $N$, where $N$ is the doping of the p or n region. Therefore, our initial hypothesis related to low-injection operation conditions is fulfilled. When the additional shading losses are considered, the amount of illumination seen by the cell is effectively reduced, so the entire current-voltage curve is shifted downward, and both the short-circuit current and open-circuit voltage are reduced. Recalling the grid model outlined in section 2.2.2, the additional series resistance losses, $r_s$, are accounted for by subtracting a voltage loss from the current-voltage characteristics calculated from solving the minority charge carrier diffusion equations (see Eq. (11)). This voltage loss decreases the slope of the curve near the open-circuit voltage and also slightly reduces the short-circuit current. Therefore, when all losses are considered, the current-voltage characteristics show this change in the slope near the open-circuit voltage as well as a downward shift of the entire curve.

## 4. Conclusions

A narrow-bandgap PV cell made of InAs with Au front contacts was designed for a NF-TPV device made of millimeter-size surfaces separated by a nanosize vacuum gap. Specifically, the architecture of the InAs cell, including the emitter and base thicknesses, the doping level of the base, and the Au front contact grid parameters were optimized for maximizing NF-TPV power output. For the first time, the impact of losses due to shading by the front contacts and losses due to series resistance introduced by the front contacts and the substrate were quantified. Results showed that these additional loss mechanisms significantly affect NF-TPV performance, and that the maximum power output is a trade-off between shading losses and series resistance introduced by the front contacts. The key conclusion is that the PV cell in NF-TPV devices must be designed by taking into account not only the temperature of the radiator, but also the vacuum gap distance at which the device operates and the surface area of the device.

Shading losses were taken into account in this work via geometric optics, a reasonable approximation provided that the front contact grid parameters were larger than the thermal wavelength. For front contact parameters of the same order of magnitude as or smaller than the thermal wavelength, it will be necessary to account for radiation scattering [72]. This is left as a future research effort.

**Acknowledgments**

D.M. and M.F. acknowledge financial support from the National Science Foundation (grant no. CBET-1253577). R.V. acknowledges financial support by the French National Research Agency (ANR) under grant No. ANR-16-CE05-0013 and is thankful to the Instituto de Energía Solar at the Universidad Politécnica de Madrid (UPM) for hosting him in 2018. E.A. acknowledges a Ramón y Cajal Fellowship (RYC-2015-18539) funded by the Spanish Science, Innovation and Universities Ministry and a Young Researcher Grant funded by Universidad Politécnica de Madrid.

**Appendix A. Supplementary data**

Supplementary data to this article can be found online at [URL to be added by the publisher].

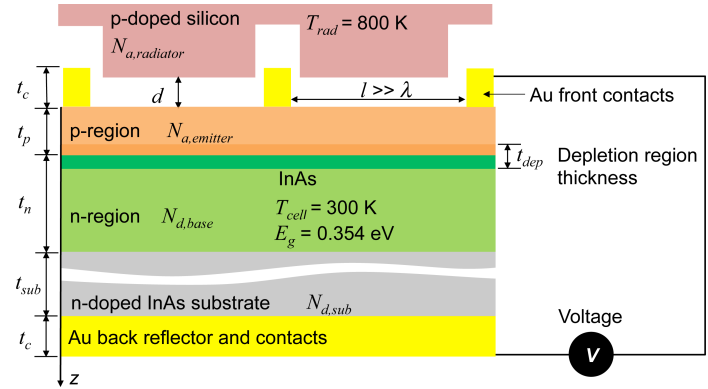

(a)

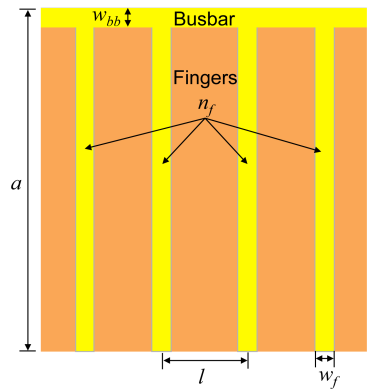

(b)

**Fig. 1.** (a) Cross-sectional view of the NF-TPV device, consisting of a p-doped Si radiator separated by a vacuum gap $d$ from an InAs PV cell. The gap $d$ is measured from the bottom surface of the Si radiator to the top surface of the InAs cell. The front and back contacts are both made of Au. (b) Top view of the metallic front contact grid consisting of a busbar and grid fingers. The dimensions of the grid fingers and their spacing are sufficiently large in comparison to the characteristic wavelength of thermal emission, $\lambda$, such that any radiation scattering by the grid is assumed to be negligible.

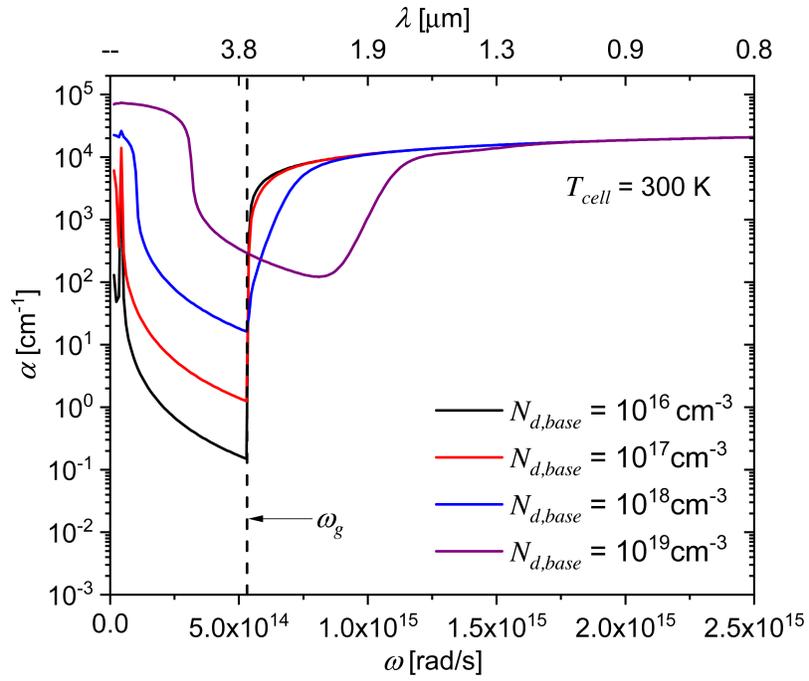

**Fig. 2.** Absorption coefficient, $\alpha$, of InAs as a function of doping concentration of the base, $N_{d,base}$. The angular frequency corresponding to the absorption bandgap of InAs ($\omega_g = 5.38 \times 10^{14}$ rad/s; $E_g = 0.354$ eV) is denoted by the vertical dashed line.

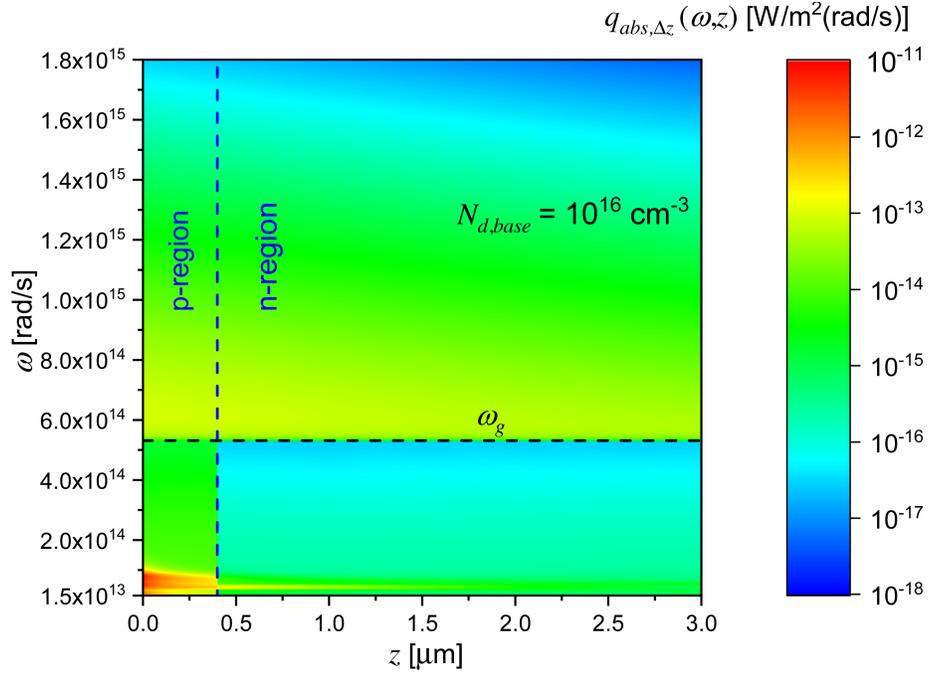

**(a)**

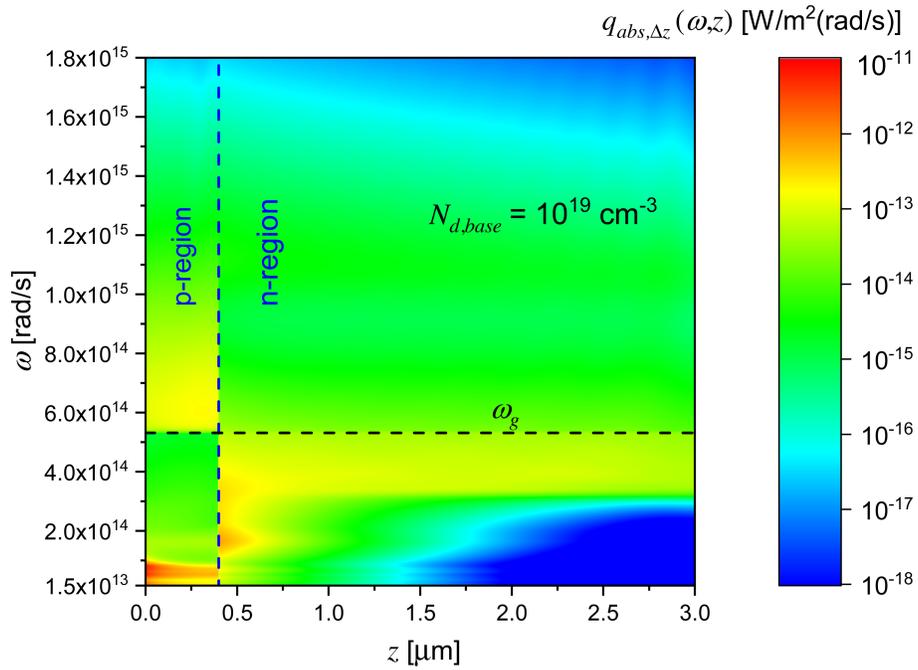

**(b)**

**Fig. 3.** Spectral-spatial absorption distribution within the InAs cell for a gap distance *d* of 100 nm: (a) $N_{d,base} = 10^{16}$ cm$^{-3}$. (b) $N_{d,base} = 10^{19}$ cm$^{-3}$. The horizontal axis represents depth within the cell, with the vertical dashed line showing the demarcation between the emitter $(0 - 0.4 \text{ } \mu m)$ and the base $(0.4 - 3 \text{ } \mu m)$ for the optimal emitter (p-region) and base (n-region) thicknesses. The horizontal dashed line shows the angular frequency ($\omega_g = 5.38 \times 10^{14}$ rad/s; $E_g = 0.354$ eV) corresponding to the bandgap of InAs.

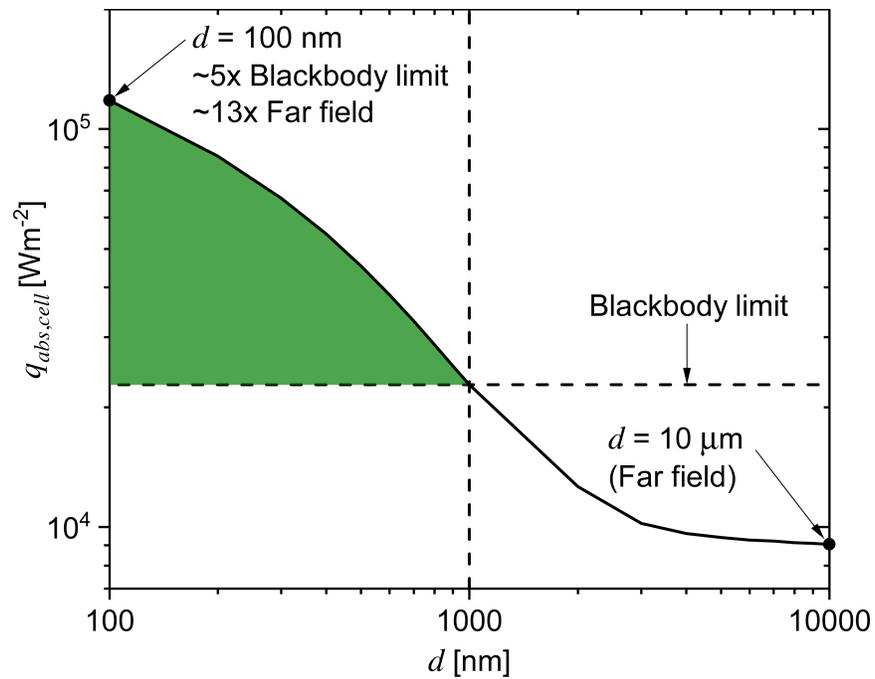

**Fig. 4.** Radiative flux absorbed by the cell (p-n junction, substrate, back contacts) as a function of vacuum gap thickness $d$ from 10 μm down to 100 nm. Shading losses due to the front contacts are not included. The blackbody limit for radiation exchange between bodies at 800 K and 300 K is indicated by the horizontal dashed line. The vertical dashed line at $d = 1$ μm is where enhancement above the blackbody limit begins to be observed.

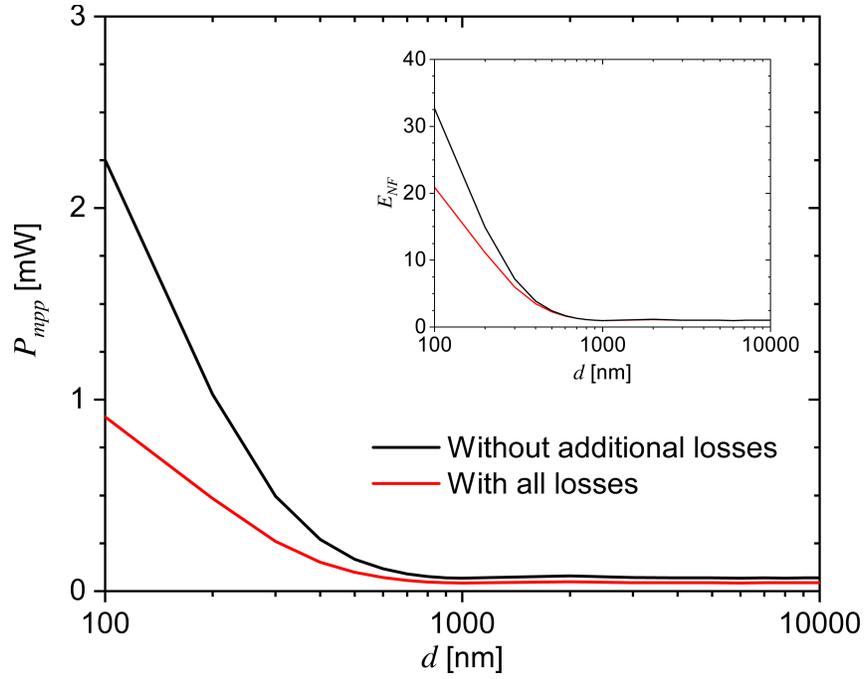

**Fig. 5.** Maximum power output $P_{mpp}$ and near-field power enhancement factor over the far-field value ($d = 10$ μm), $E_{NF}$, for a $1 \times 1$ mm² NF-TPV device as a function of the vacuum gap thickness $d$ without additional losses (due to shading and series resistance $r_s$) and with all losses. The front contact grid has been optimized for each $d$ when all losses are considered. For reference, the maximum power output with all losses and without the additional losses in the far field ($d = 10$ μm) is 0.0435 mW and 0.0690 mW, respectively.

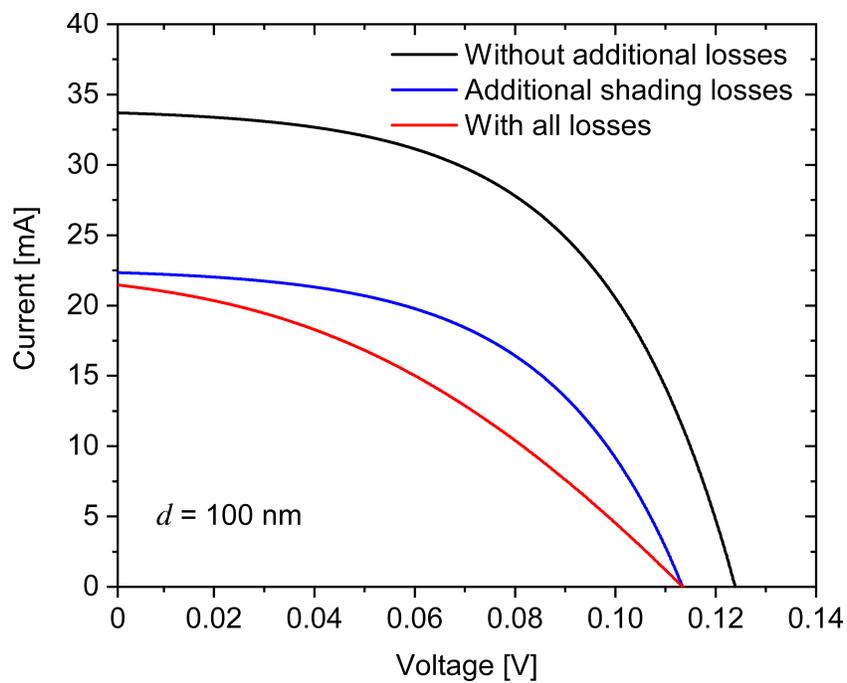

(a)

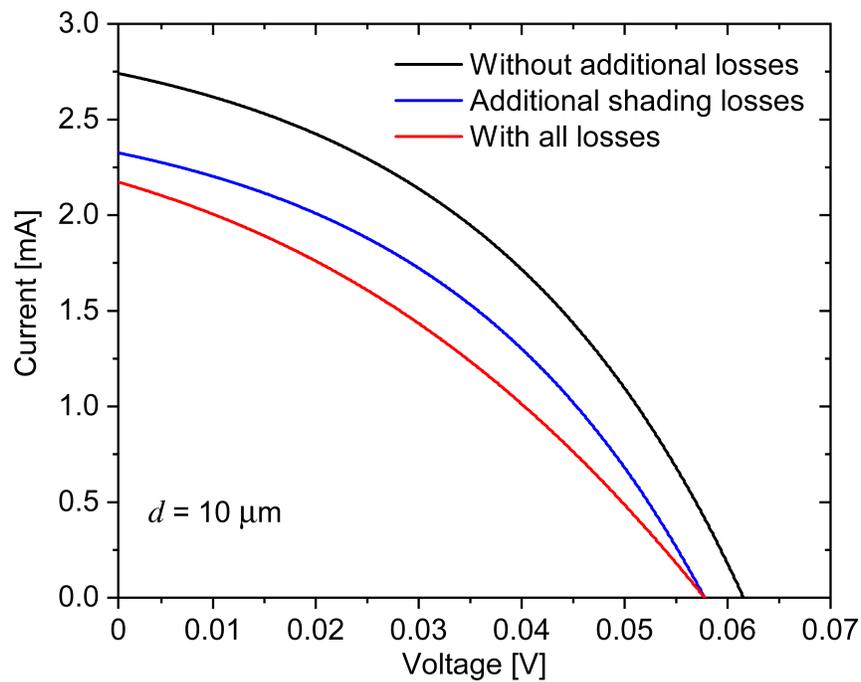

(b)

**Fig. 6.** Current-voltage characteristics for a $1 \times 1$ mm$^2$ NF-TPV device operating at a vacuum gap of (a) $d = 100$ nm, and (b) $d = 10$ μm without additional losses (due to shading and series resistance $r_s$), when only the additional shading losses are considered, and when all losses are considered. Cells are assumed to operate at 300 K.

**Table 1.** Fixed and free parameters for the InAs cell design.

| Fixed Parameters | | Free Parameters | |
|---|---|---|---|
| **Parameter** | **Value** | **Parameter** | **Range** |
| Thickness of the substrate, $t_{sub}$ | 500 µm | Thickness of the emitter, $t_p$ | 0.40 – 2.60 µm |
| Total p-n junction thickness, $t_p + t_n$ | 3 µm | Thickness of the base, $t_n$ | 0.40 – 2.60 µm |
| Thickness of the front and back contacts, $t_c$ | 200 nm | Width of the busbar, $w_{bb}$ | 20 – 1500 µm |
| Doping level of the Si radiator (p-doped), $N_{a,radiator}$ | $10^{19}$ cm⁻³ | Width of the fingers, $w_f$ | 20 – 1500 µm |
| Doping level of the substrate (n-doped), $N_{d,sub}$ | $1.5 \times 10^{17}$ cm⁻³ | Number of fingers, $n_f$ | 1 – 200 |
| Doping level of the emitter (p-doped), $N_{a,emitter}$ | $10^{18}$ cm⁻³ | Doping level of the base (n-doped), $N_{d,base}$ | $10^{16}$, $10^{17}$, $10^{18}$, $10^{19}$ cm⁻³ |
| Temperature of the radiator, $T_{rad}$ | 800 K | Surface area of the cell, $A_{cell}$ | $1 \times 1$ mm², $5 \times 5$ mm² |
| Temperature of the cell, $T_{cell}$ | 300 K | | |
| Bandgap of InAs, $E_g$ (300 K) | 0.354 eV ($\omega_g = 5.38 \times 10^{14}$ rad/s) | | |

**Table 2.** Parameters used for calculating the optical properties of the radiator and cell needed for the radiation transport model. In cases where only one reference is given, all values come from the same reference.

| Material | Parameter | Model | Inputs |
|---|---|---|---|
| p-doped Si<br>$N_{a,radiator} = 10^{19}$ cm$^{-3}$<br>$T_{rad} = 800$ K | $\varepsilon(\omega, T_{rad}, N_{a,radiator})$ | Drude [53] | $\varepsilon_\infty = 11.7$<br>$\omega_p = 2.87 \times 10^{14}$ rad/s<br>$\Gamma = 1.47 \times 10^{13}$ rad/s |
| Au | $\varepsilon(\omega)$ | Drude [54] | $\varepsilon_\infty = 1$<br>$\omega_p = 1.37 \times 10^{16}$ rad/s<br>$\Gamma = 5.32 \times 10^{13}$ rad/s |
| p-type InAs (emitter) | $\varepsilon_{FCL}(\omega, N_{a,emitter})$ | Drude-Lorentz | $\varepsilon_\infty = 12.3$ [56]<br>$\omega_p = \frac{N_{a,emitter}e^2}{\varepsilon_0 \varepsilon_\infty m_h^*}$ [43]<br>$\Gamma = \frac{e}{m_h^* \mu}$ [43]<br>$\gamma = 2.89 \times 10^{11}$ rad/s [55]<br>$\omega_{TO} = 4.12 \times 10^{13}$ rad/s [55]<br>$\omega_{LO} = 4.58 \times 10^{13}$ rad/s [52]<br>$m_{hh}^* = 0.41\ m_0$ [56]<br>$m_{lh}^* = 0.026\ m_0$ [56]<br>$m_h^* = 0.4144\ m_0$ [56,57]† |
| n-type InAs (base) | $\varepsilon_{FCL}(\omega, N_{d,base})$ | Drude-Lorentz | $\varepsilon_\infty = 12.3$ [56]<br>$\omega_p = \frac{N_{d,base}e^2}{\varepsilon_0 \varepsilon_\infty m_e^*}$ [43]<br>$\Gamma = \frac{e}{m_e^* \mu}$ [43]<br>$\gamma = 2.89 \times 10^{11}$ rad/s [55]<br>$\omega_{TO} = 4.12 \times 10^{13}$ rad/s [55]<br>$\omega_{LO} = 4.58 \times 10^{13}$ rad/s [55]<br>$m_e^* = 0.026\ m_0$ [47] |
| n-type InAs (substrate) | $\varepsilon_{FCL}(\omega, N_{d,sub})$ | Drude-Lorentz | $\varepsilon_\infty = 12.3$ [56]<br>$\omega_p = \frac{N_{d,sub}e^2}{\varepsilon_0 \varepsilon_\infty m_e^*}$ [43]<br>$\Gamma = \frac{e}{m_e^* \mu}$ [43]<br>$\gamma = 2.89 \times 10^{11}$ rad/s [55] |
| | | | $\omega_{TO} = 4.12 \times 10^{13}$ rad/s [55] |
| | | | $\omega_{LO} = 4.58 \times 10^{13}$ rad/s [55] |
| | | | $m_e^* = 0.026 \, m_0$ [47] |
| InAs (emitter/base/substrate) | $\varepsilon_{IB}(\omega, T)$ | Ref. [61] | $E_g = 0.417 - \frac{0.276 \, 10^{-3} T^2}{(T+93)}$ eV [47] |
| | | | $\bar{\alpha} = \frac{1}{E_g}$ (2.857 at 300 K) eV⁻¹ [58] |
| | | | $\varepsilon_\infty = 12.3$ [53] |
| | | | $P = 8.58 \times 10^{-8}$ eV cm⁻¹ [59] |
| | | | $m_0 = 0.51099906 \times 10^6$ eV $c_0^{-2 \, \ddagger}$ |
| | | | $m_{hh} = 0.41 \, m_0$ [56] |
| | | | $m_e = 0.026 \, m_0$ [47] |
| | | | $k_B = 8.617385 \times 10^{-5}$ eV K⁻¹ |
| | | | $T = 300$ K |

† Density-of-state effective mass of the valence band, calculated using the formulation in Ref. [57]

‡ $c_0$ is in units of cm s⁻¹

**Table 3.** Parameters used for calculating the electrical properties of the cell needed for the electrical transport model. In cases where only one reference is given, all values come from the same reference.

| Material | Parameter | Model | Inputs |
|---|---|---|---|
| InAs (emitter/base/substrate) | $\mu(N,T)$ | Caughey-Thomas [65] | $\mu_{max,e}$ = 34,000 cm$^{-2}$ V$^{-1}$ s$^{-1}$ <br> $\mu_{min,e}$ = 1000 cm$^{-2}$ V$^{-1}$ s$^{-1}$ <br> $N_{ref,e}$ = 1.1×10$^{18}$ cm$^{-3}$ <br> $\phi_e$ = 0.32 <br> $\theta_{1,e}$ = 1.57 <br> $\theta_{2,e}$ = 3.0 <br><br> $\mu_{max,h}$ = 530 cm$^{-2}$ V$^{-1}$ s$^{-1}$ <br> $\mu_{min,h}$ = 20 cm$^{-2}$ V$^{-1}$ s$^{-1}$ <br> $N_{ref,h}$ = 1.1×10$^{17}$ cm$^{-3}$ <br> $\phi_h$ = 0.46 <br> $\theta_{1,h}$ = 2.3 <br> $\theta_{2,h}$ = 3.0 |
| InAs (emitter/base) | $\tau_{rad}(N)$ | $\tau_{rad} = \frac{\phi}{BN}$ [56] | $\phi = 1$ <br> $B$ = 1.1×10$^{-10}$ cm$^{-3}$ s$^{-1}$ |
| InAs (emitter/base) | $\tau_{Auger}(N)$ | $\tau_{Auger} = \frac{1}{CN^2}$ [56] | $C$ = 2.2×10$^{-27}$ cm$^6$ s$^{-1}$ |
| InAs (emitter/base) | $\tau_{SRH,emitter}$ <br> $\tau_{SRH,base}$ | N/A | $\tau_{SRH,emitter}$ = 3×10$^{-9}$ s [56] <br> $\tau_{SRH,base}$ = 3×10$^{-7}$ s [56] |
| InAs (emitter/base) | $S_{emitter}$ <br> $S_{base}$ | N/A | $S_{emitter}$ = 10$^3$ cm s$^{-1}$ [56] <br> $S_{base}$ = 10 cm s$^{-1}$ [56] |

**Table 4.** Front contact grid parameters for the $1 \times 1$ mm$^2$ and $5 \times 5$ mm$^2$ NF-TPV devices with the optimized cell parameters $t_p = 0.40$ μm, $t_n = 2.60$ μm, and $N_{d,base} = 10^{16}$ cm$^{-3}$ operating at gap thicknesses of 10 μm and 100 nm along with $P_{mpp}$, $\eta_{jun}$, and $\eta_{cell}$ when all losses are considered, and without the additional losses (shading and series resistance $r_s$).

| Grid Parameters | $d = 10$ μm | | $d = 100$ nm | | |
|---|---|---|---|---|---|
| | $1 \times 1$ mm$^2$ (optimized) | $5 \times 5$ mm$^2$ (optimized) | $1 \times 1$ mm$^2$ (optimized) | $5 \times 5$ mm$^2$ (optimized) | $5 \times 5$ mm$^2$ (scaled-up from optimized $1 \times 1$ mm$^2$ grid) |
| $w_{bb}$ [μm] | 31 | 569 | 77 | 959 | 76 |
| $n_f$ | 6 | 59 | 13 | 99 | 65 |
| $w_f$ [μm] | 20 | 20 | 20 | 20 | 20 |
| $l$ [μm] | 167 | 84.7 | 76.9 | 50.5 | 76.9 |
| $P_{mpp}$ (without additional losses) [mW] | 0.0690 | 1.72 | 2.25 | 56.3 | 56.3 |
| $P_{mpp}$ (with all losses) [mW] | 0.0435 | 0.397 | 0.910 | 3.53 | 0.982 |
| $\eta_{jun}$ (without additional losses) [%] | 5.10 | 5.08 | 9.67 | 9.68 | 9.68 |
| $\eta_{jun}$ (with all losses) [%] | 3.22 | 1.17 | 3.91 | 0.607 | 0.169 |
| $\eta_{cell}$ (without additional losses) [%] | 0.763 | 0.761 | 1.91 | 1.91 | 1.91 |
| $\eta_{cell}$ (with all losses) [%] | 0.481 | 0.176 | 0.771 | 0.120 | 0.033 |

**Supplementary Data**

**Design of an indium arsenide cell**

**for near-field thermophotovoltaic devices**

Daniel Milovich[a], Juan Villa[2], Elisa Antolin[b], Alejandro Datas[b], Antonio Marti[b],

Rodolphe Vaillon[c,d,b], Mathieu Francoeur[a,†]

[a]Radiative Energy Transfer Lab, Department of Mechanical Engineering, University of

Utah, Salt Lake City, UT 84112, USA

[b]Instituto de Energía Solar, Universidad Politécnica de Madrid, 28040 Madrid, Spain

[c]IES, Univ Montpellier, CNRS, Montpellier, France

[d]Univ Lyon, CNRS, INSA-Lyon, Université Claude Bernard Lyon 1, CETHIL

UMR5008, F-69621, Villeurbanne, France

† Corresponding author.
*Email address:* mfrancoeur@mech.utah.edu

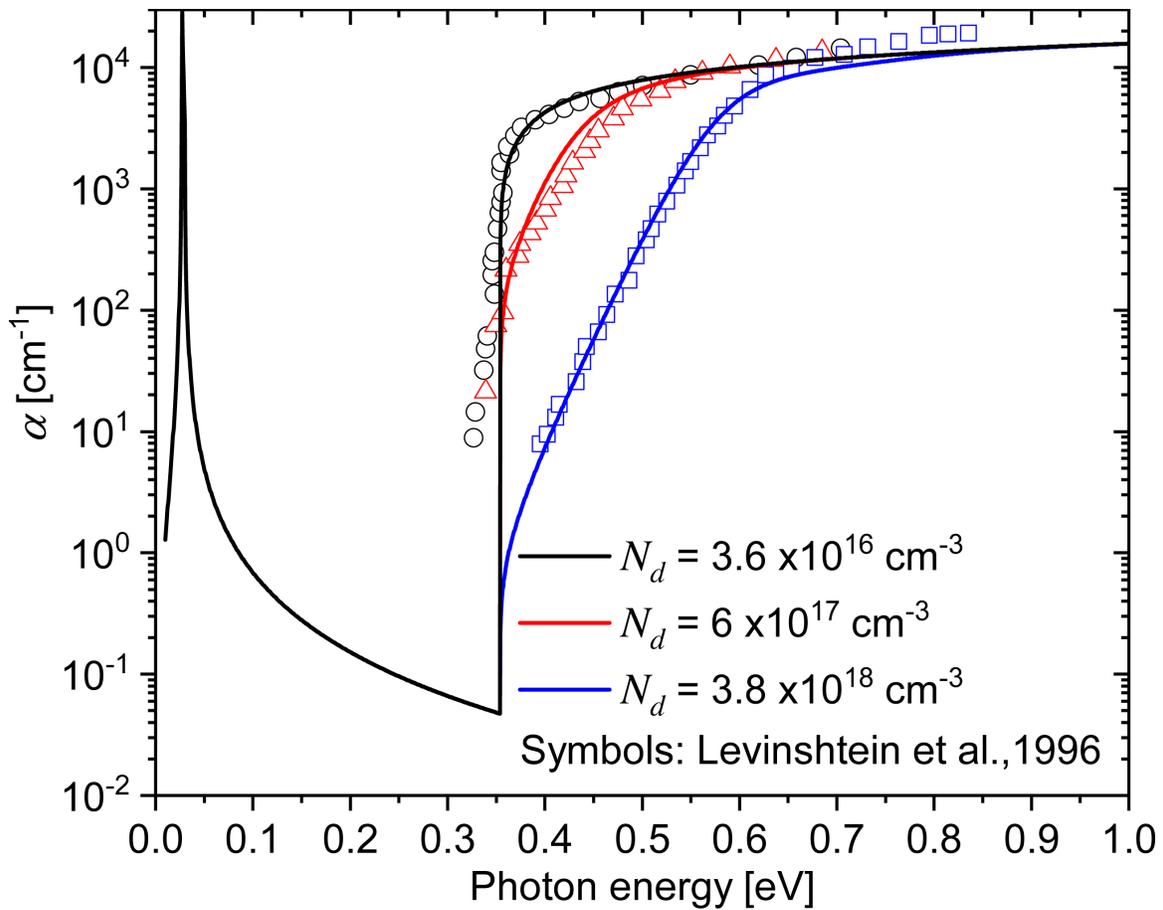

**Fig. S1.** Absorption coefficient, $\alpha$, for n-doped InAs calculated using the model described in section 2.2.1. The calculations are compared against the experimental data provided in Ref. [1]. For photon energy below the cell bandgap (0.354 eV), all curves result in the same absorption coefficient.

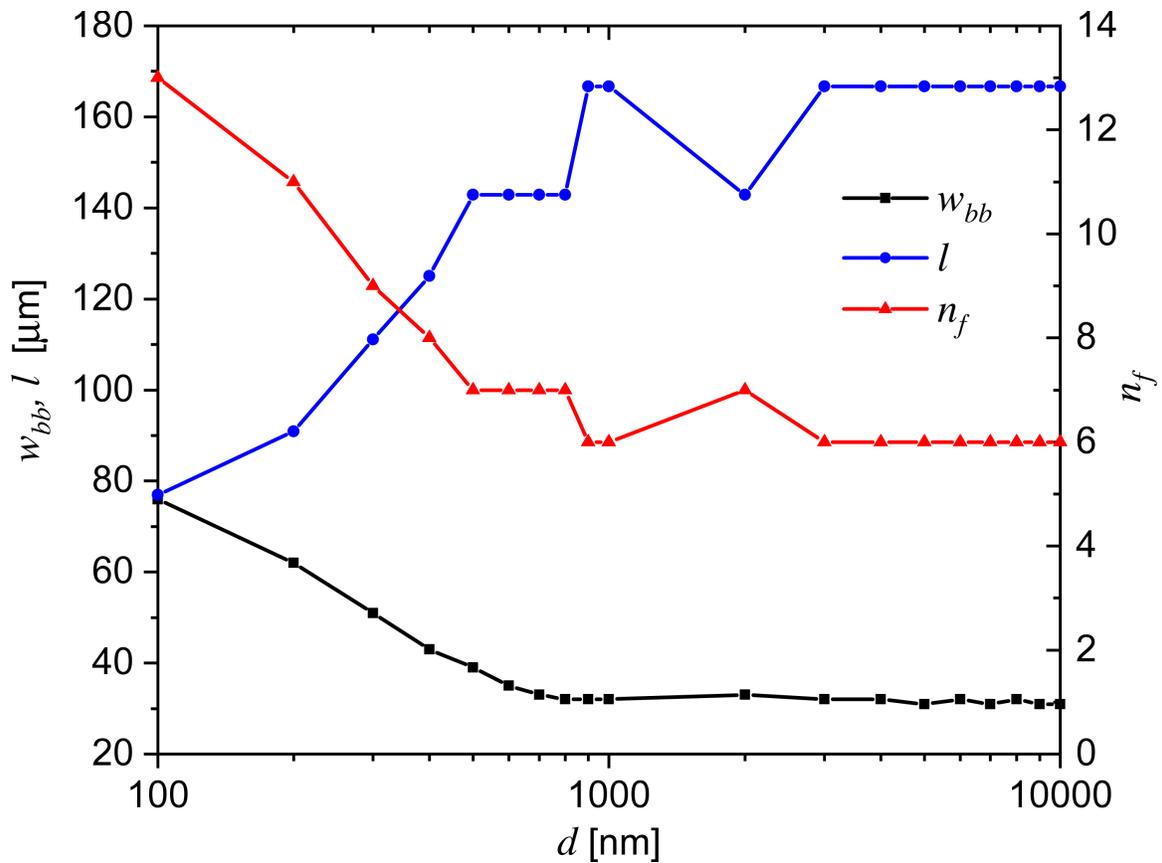

**Fig. S2.** Optimized front contact grid parameters (width of the busbar, $w_{bb}$, spacing between grid fingers, $l$, and number of fingers, $n_f$) for a $1 \times 1$ mm² NF-TPV device as a function of vacuum gap when all losses are considered. The optimal value for the width of the fingers, $w_f$, is 20 μm for all vacuum gaps. The number of fingers and their spacing are linked due to the assumption that the fingers are equally spaced.